\documentclass{article}
\usepackage{graphicx}
\usepackage{array}
\usepackage{amssymb}
\usepackage{amsfonts}
\usepackage{amsmath}
\usepackage{mathrsfs}
\usepackage{color}
\newcommand{\EqLabel}[1]{\label{#1}}
\newcommand{\Eqref}[1]{Eq. (\ref{#1})}
 
\def\ie{{i.e}. }

\begin{document}

\title{Stability of Mixed-Strategy-Based Iterative Logit Quantal Response Dynamics in Game Theory}
\author{Qian Zhuang$^{1,2}$, Zegnru Di$^{1}$, Jinshan Wu $^{1,\dag}$
\\ 1. Department of Systems Science, School of Management,
\\ Beijing Normal University, Beijing, 100875, P.R. China 
\\ 2. College of Information Science and Technology, 
\\ Nanjing Agricultural University, Nanjing, 210095, P.R. China}

\maketitle

\begin{abstract}
Using the Logit quantal response form as the response function in each step, the original definition of static quantal response equilibrium (QRE) is extended into an iterative evolution process. QREs remain as the fixed points of the dynamic process. However, depending on whether such fixed points are the long-term solutions of the dynamic process, they can be classified into stable (SQREs) and unstable (USQREs) equilibriums. This extension resembles the extension from static Nash equilibriums (NEs) to evolutionary stable solutions in the framework of evolutionary game theory. The relation between SQREs and other solution concepts of games, including NEs and QREs, is discussed. Using experimental data from other published papers, we perform a preliminary comparison between SQREs, NEs, QREs and the observed behavioral outcomes of those experiments. For certain games, we determine that SQREs have better predictive power than QREs and NEs.
\end{abstract}

\tableofcontents

\section{Introduction}

Game theory has become a powerful and popular tool in many sociological studies. Although several studies have questioned predictive power of the Nash equilibrium (NE) \cite{Holt:critical_NE,Ochs:MNE}, it has been used as a primary game solution since its initial proposition \cite{Nash:PNAS, Nash:NE}. However, the questions of finding such NEs and refining them when multiple NEs exist are not easy tasks  \cite{Selton:Selection, Samuelson:Selection}. Furthermore, people are interested to know how, in experiments or real-life observations, one ``preferred'' NE emerges from all possible strategy profiles, particularly in a population that does not begin with a NE as the initial strategic state. This phenomenon is the well-known question of learning in games and the converging towards particular solutions \cite{Fudenberg:Learning_In_Games, Fudenberg:Learning_and_Equilibrium}.

To find game solutions with predictive power, in addition to first searching for all NEs and then refining them   \cite{Myerson:refining}, dynamic processes have been proposed to describe, mimic or reproduce to a certain extent the strategic thinking processes of game players in the hope that certain long-term solutions of the dynamic processes will lead to the ``preferred'' NEs  \cite{Samuelson:Selection,Sabourian:Selection}. Well-known examples of such dynamic process include replicator dynamics   \cite{Weibull:Evolution,Hofbauer:Replicator,Boorgers:RD}, Logit learning  \cite{Alos:Logit}, and fictitious play   \cite{Brown:Fictitious,Berger:Fictitious}. In certain cases, a refined NE fits the experimental data well. We refer to such an NE as the preferred NE. In this case, a proposed evolutionary model is a good theory if the model predicts that long-term solutions of the corresponding dynamic processes converge to the refined NE. Alternatively, in other cases, no NE can explain the observed behavior in real experiments. In this case, a good theory means that long-term solutions of the proposed dynamic processes can explain the observed behavior instead of the NEs   \cite{Holt:Contradiction}. To simplify our terminology, we denote both the NE and the long-term solution in these cases, where they are capable of describing experimental or real-life observations, the preferred NE. The primary goal of these typical dynamic processes, and thus of all of these theories, is to determine the preferred NE by solving for the long-term solutions. For a dynamic process, usually two central topics should be discussed: how well experimental observations can be explained by long-term solutions of the dynamic process and the relation between the dynamic process's long-term solutions and other solution concepts such as NEs and refined NEs.

In this manuscript, we study properties of a new dynamic process: the iterative Logit quantal response dynamics (ILQRD), which will be defined based on the concept of static Quantal Response Equilibriums (QREs). Our goal of proposing this new dynamic process is solely to capture the preferred NE with long-term stable solutions of ILQRD, which we denote as stable QREs (SQREs).

This manuscript is organized as follows. In this introduction, we first explain our main idea: the evolutionary process. In section $\S$ \ref{sec:notation}, we define several notations and the dynamic process. There, we also compare the new dynamic process with other learning models and evolutionary processes in game theory. In the rest parts of this manuscript, we attempt to discuss the two previously introduced central topics of this ILQRD process: how the long-term solutions fit experimental results and what is the relation between its long-term solution and other solution concepts. In section $\S$ \ref{sec:example}, we illustrate the performance of our dynamic process using examples. In section $\S$ \ref{sec:proof}, we provide an analytical proof of the major conclusions for the special cases of $2\times2$ symmetrical games. Then, in section $\S$\ref{sec:QRSS}, we compare our SQRE with a highly similar solution concept: the quantal response stable solutions (QRSS). In section $\S$ \ref{sec:exp}, using SQRE, we re-analyze some collected experimental results. Finally, in section $\S$\ref{sec:conclusion}, we summarize our main conclusions and discuss possible future recearch.

\section{Notations and Definitions}
\label{sec:notation}

To present our formula in a compact form and to remain consistent with the notations and the terminology of statistical ensembles used in statistical physics and quantum statistical physics, in this section, we introduce a matrix-based notation to represent a general $N\times M$ game. The key notation that differs from the conventional mathematical forms of game theory is the matrix representation of probability distributions and the payoff matrices of general $N$-player games. One may proceed directly to \Eqref{eq:iteration21} and \Eqref{eq:iteration12} and continue from there if learning these new notation presents an obstacle. Most of our expressions can be understood in terms of the conventional mathematics of game theory. However, we believe that they can be understood more conveniently using the new notation. Furthermore, the new notation is readily applicable to quantum games   \cite{Jinshan:Qgame}.

\subsection{A new set of matrix-based notations}
Here, we introduce a matrix-based notation for probability distributions such that the probability distribution of the strategic status of all players and the mathematical description of payoffs for games with an arbitrary number of players and an arbitrary number of strategies become matrices. However, to understand our work in this manuscript, this new set of notation is not necessary. One may skip this section and proceed directly to \Eqref{eq:iteration21} and \Eqref{eq:iteration12} in $\S$\ref{subsec:model}.

Consider a $2\times2$ game with the following conventional form of payoff bi-matrix:
\begin{align}
G^{1,2} = \left[\begin{array}{cc} a, a^{'} & b, c^{'} \\ c, b^{'} & d, d^{'}\end{array}\right],
\EqLabel{eq:payoffmatrix}
\end{align}
with the convention that the row (column) strategies belong to the first (second) player and the first (second) number of all entries represents the payoff received by the first (second) player. We denote the first (second) player's strategy $C,D$ (also $C,D$, although one player's strategies can be totally different from the other player's strategies). Usually, mixed strategies, which include pure strategies as special cases, are written as column vectors: For example,
\begin{align}
P^{1} = \left[p^{1}, 1-p^{1}\right]^{T}
\EqLabel{eq:p1}
\end{align}
for player $1$ and
\begin{align}
P^{2} = \left[p^{2}, 1-p^{2}\right]^{T}
\EqLabel{eq:p2}
\end{align}
for player $2$.
The payoff is calculated from the following vector-matrix-vector multiplication£º
\begin{align}
E^{1,2} = \left(P^{1}\right)^{T}G^{1,2}P^{2}.
\EqLabel{eq:oldpayoff}
\end{align}
For $2\times2$ games, the payoff matrix $G$ indeed appears as a matrix. However, for a general $N\times M$ game, the matrix becomes a map from $N$ vectors to $\mathcal{R}$, \ie, cubic tensors for $3$-player games and $T\left(N,0\right)$-type tensors for $N$-player games. The payoff is no longer a matrix.

To unify the notation of all $N\times M$ games, we introduce a new equivalent set of notations as follows. We write payoff matrices and mixed strategies as matrices:
\begin{align}
H^{1,2} = \left[\begin{array}{cccc} a, a^{'} & 0 & 0 & 0 \\ 0 & b, c^{'} & 0 & 0 \\ 0 & 0 & c, b^{'} & 0 \\ 0 & 0 & 0 & d, d^{'}\end{array}\right],
\end{align}
and
\begin{align}
\rho^{1,2} = \left[\begin{array}{cc} p^{1}, p^{2} & 0 \\ 0 & \left(1-p^{1}\right), \left(1-p^{2}\right) \end{array}\right].
\end{align}
Then we calculate the payoff from the following trace operation:
\begin{align}
E^{1,2} = tr\left(H^{1,2}\rho^{1}\otimes\rho^{2}\right).
\EqLabel{eq:newpayoff}
\end{align}
One can confirm that with the same strategy profiles, \Eqref{eq:newpayoff} and \Eqref{eq:oldpayoff} result in the same payoffs. For the special case of the $2\times 2$ game, in the new formalism, the payoff matrices and the matrices of strategy state of all players are of dimension $2^{2}$. The probability distribution of both players is defined as a direct product of each player's state matrix:
\begin{align}
\rho = \rho^{1}\otimes\rho^{2}.
\end{align}
Every entry of this state matrix corresponds to the probability of all of the players choosing the corresponding strategic combination defined by the position of the entry. For example, the $\left(1,1\right)$ (upper-left) entry of $\rho$ is $p^{1}p^{2}$ and means that at this probability, the two players take the strategic combination $\left(C,C\right)$. In turn, the correspondence of this entry in the payoff matrices $H^{1,2}$ -- their $\left(1,1\right)$ entries -- are naturally $\left(a,a^{'}\right)$. In this sense, from the general expression,
\begin{align}
E^{i} = tr\left(H^{i}\rho^{1}\otimes\rho^{2}\right)\equiv H\left(\rho^{1}, \rho^{2}\right).
\EqLabel{eq:Hmap}
\end{align}
$H^{i}$ can be interpreted as a linear map from the set of $\left\{\left(\rho^{1}, \rho^{2}\right)\right\}$ to real number $\mathcal{R}$.

Another useful feature of this notation system is that it streamlines the description of correlated strategies. That is, this notation also functions when $\rho \neq \rho^{1}\otimes\rho^{2}$. For an $N\times M$ game, $s^{i}_{l}$ stands for the $l$th strategy of the $i$th player, where $i\in [1,N]$ and $l\in [1,M]$. The set of strategies of player $i$ is denoted as $S^{i}=\left\{s^{i}_{1}, s^{i}_{2}, \cdots, s^{i}_{M}\right\}$. For convenience, we denote the set of probability distributions over $S^{i}$ as $\Delta^{i}$, \ie, $\Delta^{i} = \left\{\rho^{i}\right\}$ and the direct product set of all of these probability distributions as $\Delta$, \ie, $\Delta = \Delta^{1}\otimes\Delta^{2}$. This $\Delta$ differs from the set of probability distributions over $S=S^{1}\otimes S^{1}$, which we denote as $\Delta\left(S\right)$. This space includes the correlated strategy, whereas $\Delta$ is the set of only independent strategies. For example, in this notation, a general possibly correlated equilibrium   \cite{Aumann:CE} can be defined as $\rho^{12}_{ce} \in \Delta\left(S\right)$ such that for every player $i$, $\forall \rho^{i} \in \Delta^{i}$,
\begin{align}
tr\left(H^{i}\rho^{1,2}_{ce}\right) \ge tr\left(H^{i}\rho^{i}\otimes tr^{i}\left(\rho^{12}_{ce}\right)\right)
\EqLabel{eq:CE}
\end{align}
where $tr^{i}\left(\rho^{1,2}_{ce}\right)$ is a partial trace, which performs the partial integral/summation over player $i$'s strategy space. For example,
\begin{align}
tr^{1}\left(\rho^{12} \right) =\sum_{l} \rho^{12} \left(s^{1}_{l}, \cdot\right)
\EqLabel{eq:tr}
\end{align}
and the result is a strategy profile of player $2$. This partial trace is the same as the partial summation in deriving partial distribution in probability theory. In our notation, NE is defined as $\rho^{1}_{ne} \otimes \rho^{2}_{ne}$ or, equivalently, $\rho^{12}_{ne} \in \Delta$ such that for $ \forall i, \forall \rho^{i}$,
\begin{align}
tr\left(H^{i}\rho^{1}_{ne} \otimes \rho^{2}_{ne}\right) \ge tr\left(H^{i}\rho^{i}\otimes tr^{i}\left(\rho^{1}_{ne} \otimes \rho^{2}_{ne}\right)\right).
\EqLabel{eq:NE}
\end{align}

\subsection{Definition of iterative Logit quantal response dynamics (ILQRD) and its stable equilibriums}
\label{subsec:model}
Using the previously described matrix-based notation, for a $2$-player game our iterative Logit quantal response dynamics (ILQRD) is defined as follows:
\begin{align}
\rho^{1}\left(t+1\right) = \frac{1}{Z_{R}^{1}\left(t\right)}e^{\beta H^{1}_{R}\left(\rho^{2}\left(t\right)\right)}, \EqLabel{eq:map2to1}\\
\rho^{2}\left(t+1\right) = \frac{1}{Z_{R}^{2}\left(t+1\right)}e^{\beta H^{2}_{R}\left(\rho^{1}\left(t+1\right)\right)} \EqLabel{eq:map1to2},
\end{align}
where the reduced payoff matrix is defined as
\begin{align}
 H^{1}_{R}\left(\rho^{2}\right) =H^{1}\left(\cdot, \rho^{2}\right), \\
 H^{2}_{R}\left(\rho^{1}\right) =H^{2}\left( \rho^{1},\cdot\right).
\end{align}
Normalization constant $Z^{i}_{R}$ is defined as
\begin{align}
Z_{R}^{i} = tr\left(e^{\beta H^{i}_{R}}\right).
\end{align}
Using the matrix-based notations, one can straightforwardly extend this ILQRD to general $N\times M$ games.

Using the usual notation of probability distribution, for the $2\times 2$ game defined in \Eqref{eq:payoffmatrix}, \Eqref{eq:p1} and \Eqref{eq:p2}, we can rewrite the iteration process explicitly as follows:
\begin{align}
 p^{1}(t+1) = \frac{e^{\beta\left(ap^{2}(t)+b(1-p^{2}(t))\right)}}{e^{\beta\left(ap^{2}(t)+b(1-p^{2}(t))\right)}+e^{\beta\left(cp^{2}(t)+d(1-p^{2}(t))\right)}}, \EqLabel{eq:iteration21} \\
p^{2}(t+1) = \frac{e^{\beta\left(a^{'}p^{1}(t+1)+b^{'}(1-p^{2}(t+1))\right)}}{e^{\beta\left(a^{'}p^{1}(t+1)+b^{'}(1-p^{1}(t+1))\right)}+e^{\beta\left(c^{'}p^{1}(t+1)+d^{'}(1-p^{1}(t+1))\right)}}
\EqLabel{eq:iteration12}
\end{align}
To simplify our notation, we denote the RHS of \Eqref{eq:iteration21} as $g\left(p; \beta \left|\right. a, b, c, d\right)$, where $a,b,c$ and $d$ can be omitted when it is clear what the parameter $a,b,c$ and $d$ refers to. Formally, we denote this map as
\begin{align}
p^{2}\left(t+1\right) = \phi\left(p^{2}(t)\right) = g\left(g\left(p^{2}(t); \beta \left|\right. a, b, c, d\right); \beta \left|\right. a^{'}, b^{'}, c^{'}, d^{'}\right)
\end{align}
and $p^{1}\left(t+1\right)$ can be regarded as an intermediate variable. This map is an iterative map from a mixed strategy ($p^{2}(t)$) to a new mixed strategy ($p^{2}(t+1)$).

The fixed points of this ILQRD are the same as the static Logit QRE, and they are denoted as $\rho_{*}\left(\beta\right)=\left(p_{*}^{1}\left(\beta\right), p_{*}^{2}\left(\beta\right)\right)$. If a fixed point of those QREs is also the long-term evolution of ILQRD, this fixed point is referred to as a stable QRE (SQRE) and denoted as $\rho_{\infty}\left(\beta\right)=\left(p^{1}_{\infty}\left(\beta\right), p^{2}_{\infty}\left(\beta\right)\right)$. Otherwise, it will be referred to as an unstable QRE (USQRE). Next, we focus on the relations among pure-strategy NEs, mixed NEs, QREs and SQREs and experimental data. A SQRE must be a QRE. However, the inverse is not necessary true. This stability test potentially differentiates QREs into SQREs and USQREs. In principle, such differentiation can improve the predictive power of QREs for experimental data and examing/demonstrating this is the whole point of the present manuscript.

\subsection{Difference between our evolutionary process and other learning/imitating models}
In ILQRD, a key concept is the use of the quantal response function (as in \Eqref{eq:map2to1} and \Eqref{eq:map1to2}) to determine a player's strategy profile according to the player's corresponding payoffs. The introduction of parameter $\beta$ as a description of bounded rationality in this form of quantal response function is common in theories of learning in games   \cite{Chen:BoundedNE} and in the QRE concept \cite{McKelvey:QRE1}. Additionally, this idea may be proposed simply from the viewpoint of statistical physics   \cite{Jinshan:Qgame}. The use of parameter $\beta$ can be justified to a certain degree based on games with limited information   \cite{Wolpert:Infor, Haile:QRE2}.

In fact, the same expression used in \Eqref{eq:map2to1} and \Eqref{eq:map1to2} has been used in discussions of Logit QRE, and a highly similar expression has been used in Logit learning \cite{Baron:QRSS, Alos:Logit}, stochastic fictitious play \cite{Brown:Fictitious} and stochastic reinforcement learning \cite{Hopkins:TwoModels}. In these theories, the quantal response function, as in \Eqref{eq:map2to1} and \Eqref{eq:map1to2}, is occasionally referred to as the smoothed best response. However, all of these theories differ from ours in principle, as explained bellow.

First, we compare our expressions with the QRE. In the QRE,
\begin{align}
p\left(s^{i}\right) = \frac{e^{\beta u^{\left(i\right)}\left(s^{i},p^{-i}\right)}}{\sum_{r^{i}\in S^{i}}e^{\beta u^{\left(i\right)}\left(r^{i},p^{-i}\right)}}
\EqLabel{eq:QRE}
\end{align}
is a map from all players' strategy profile $p=\Pi \otimes p^{i}$ to itself. This equation is a fixed-point equation. Our work differs from the QRE in that a QRE only focuses on fixed points solved from static equations, whereas we use iterations to find the stable fixed points and distinguish them from other, unstable fixed points. Later, we will note that such a difference in stability is essential in applying the QRE to explaining experiments. 

The QRE has been compared with experimental observation and generally provides a better fit to the data than a NE   \cite{McKelvey:QRE_NE}. However, the QRE has been criticized as an illusory improvement because there is an additional free parameter in QRE when fitting the curve, and one can always improve using an additional parameter. We demonstrate that this free parameter is not completely free. In fact, in certain cases, when the $\beta$ is sufficiently large, the fixed points from the QRE are no longer stable. Therefore, when comparing experimental data with the stable/unstable QREs and the NEs, one can determine whether the QRE or the NE has more predictive power. Distinguishing stable QREs from unstable QREs using iterative dynamics is this manuscript's first contribution. As presented below, we collected experimental data and conducted a preliminary comparison of the theories with experimental observations. After distinguishing SQREs from QREs, we tested SQREs, QREs and NEs against several experiments reported in the literature, which is this manuscript's second contribution. One possible further investigation along this line, which has not been implimented in this work, can be applying our ILQRD to cross-game experiments. For example, using the same players in different games with similar level of payoffs, we can estimate the parameter $\beta$ from one game and test it in other games.

In a dynamic QRE \cite{Baron:QRSS}, so-called Logit learning \cite{Alos:Logit}, which is driven by observations during real game-playing processes, in which each player chooses only one pure strategy to play every turn, the same smoothed best response function is used to mimic the player's response to the opponent's pure strategy, as follows:
\begin{align}
p\left(s^{i},s^{-i}\right) = \frac{e^{\beta u^{\left(i\right)}\left(s^{i},s^{-i}\right)}}{\sum_{r^{i}\in S^{i}}e^{\beta u^{\left(i\right)}\left(r^{i},s^{-i}\right)}}.
\EqLabel{eq:QRSS}
\end{align}
In fact, this smoothed best response function defines a probability transition matrix between the current strategy profiles of player $i$ and the previous strategy profiles of all of the other players. This transition probability depends not on player $i$'s previous state $s^{i}\left(t-1\right)$ but on the previous states of all of the other players $s^{-i}\left(t-1\right)$. For simplicity, we express this probability as follows:
\begin{align}
M^{\left(i\right)}\left(s^{-i}\left(t\right)\right)=\left[p\left(s^{i},s^{-i}\left(t\right)\right)\right]_{M\times M}.
\end{align}
If we let each player take his or her turn in the natural order, we will have a transition matrix between the current and the previous strategy profiles of all of the players. For simplicity, we denote this matrix as
\begin{align}
M=\Pi_{i=1}^{N} M^{\left(i\right)}.
\end{align}
In the abovr formula, taking $N=2$ as example, the matrix may be written according to one of the two following rules:
\begin{align}
M=M^{\left(1\right)}\left(s^{-1}\left(t-1\right)\right)M^{\left(2\right)}\left(s^{-2}\left(t\right)\right), \EqLabel{alter}\\
M=M^{\left(1\right)}\left(s^{-1}\left(t-1\right)\right)M^{\left(2\right)}\left(s^{-2}\left(t-1\right)\right). \EqLabel{simu}
\end{align}
The first rule is referred to as alternating updating, whereas the second rule is referred to as simultaneous updating. Regardless of the form assumed, the central task is to determine the invariant probability distribution using the transition matrix $M$:
\begin{align}
P_{ss}=\lim_{n\rightarrow \infty}\left(M^{T}\right)^{n}P_{0}.
\EqLabel{eq:QRSSFinal}
\end{align}
To distinguish fixed-point solutions of this transition matrix from a QRE, these long-term solutions are  occasionally referred to as end results of Logit response dynamics   \cite{Alos:Logit}. Here, we name such solutions quantal response stable solutions (QRSSs). There have been many attempts to solve   \cite{Konno:Logit_Exact} or characterize   \cite{Alos:Logit} such a QRSS ($P_{ss}$) for a given transition matrix $M$. Because a QRE and a QRSS use similar formulae, in principle, the two should be closely related. For example, a QRSS could be a subset of QREs and thus expected to be a somehow refined one from the entire set of QREs. In a sense, the stable solutions of our ILQRD also capture the stable portion of QREs. Therefore, in principle, our SQREs should be closely related to a QRSS. However, in section $\S$\ref{sec:QRSS} we demonstrate that it is generally not the case: QRSSs differ substantially from QREs and SQREs. Differentiating a QRSS from a QRE and a SQRE is this manuscript's third contribution.

Next, we focus on a comparison between ILQRD and fictitious play  \cite{Brown:Fictitious}. In fictitious play, players update their beliefs and choose a pure strategy to play according to certain decision-making rules that relate their strategy choice to their beliefs. Such decision-making rules typically include, for example, the best response myopic strategy   \cite{Brown:Fictitious} and the smoothed best response   \cite{Fudenberg:Fictitious_Smooth}. The latter uses the same expression as used in \Eqref{eq:QRE} with only one difference. The difference is, $p^{-i}$, which is the true current strategy profiles of the others, is replaced by player $i$'s belief regarding the strategy profiles of the others, which is usually taken to be the empirical distribution deduced from the entire history of other players' choices. In \Eqref{eq:QRE}, \Eqref{eq:map2to1} and \Eqref{eq:map1to2}, there is no belief and no empirical distribution. When we examine the learning process in real life, it may seem to be more reasonable to take player beliefs into consideration. However, as we have previously noted, we are substantially much concerned about finding the proper solutions, that is, solutions capable of predicting experimental behavioral outcomes, than making the entire dynamic process meaningful. Because fictitious play extracts the empirical distribution of other player strategy profiles from history, the speed of convergence occasionally becomes a problem  \cite{Jordan:Attack_Fictitious,Krishna:Attack_Fictitious}. As discussed below, in ILQRD, convergence speed is never an issue.

Similar relation holds bewteen ILQRD and stochastic reinforcement learning \cite{Hopkins:TwoModels}. A record of scores for every potential strategy is kept by every player in the reinforcement learning while here in the ILQRD, only the previous mixed strategy is used in the decision making of the current strategy.

Another widely used dynamic process to determine the preferred NE is replicator dynamics \cite{Hofbauer:Replicator, Boorgers:RD}. All such previously mentioned learning-based mapping or dynamics differ from replicator dynamics, where each player plays against a finite or infinite population and individuals learn from simple imitation but not with individual introspective thinking. In this manuscript, we focus on the effects of introspective thinking and only of the two players but not in a model of population dynamics.

We should note that the same notion of ILQRD (referred to as Boltzmann iteration) was proposed in a 2004 unpublished working paper \cite{Jinshan:Qgame} by one of the authors in a quantum game context. The idea is not a central topic in that working paper and was not developed any further there.

Additionall, a highly similar dynamic process was proposed in \cite{Holt:Introspection}, as a concept referred to as noisy rational strategies (NRS):
\begin{align}
\rho_{nrs} = \lim_{n\rightarrow\infty} \phi^{\beta_{0}}  \circ \phi^{\beta_{1}}   \circ \cdots  \circ \phi^{\beta_{n}} \left(\rho\left(0\right)\right).
\EqLabel{eq:brs}
\end{align}
There, the authors focus on the effect of increasing $\beta$ ($\beta_{0}\leq\beta_{1}\leq \dots \leq \beta_{n}$) and assume $\beta_{\infty}=\infty$ so that $\rho\left(0\right)$ becomes irrelevant. According to the authors, such an increase in $\beta$ can be interpreted as the increasing difficulty of performing a greater number of iterations given a player's finite capability   \cite{Holt:Introspection}. In this sense, what we discuss in this paper bears a greater resemblance to the following:
\begin{align}
\rho^{2}_{sqre} = \lim_{n\rightarrow\infty} \phi^{\beta_{n}}  \circ \cdots \circ \phi^{\beta_{2}}   \circ \phi^{\beta_{1}} \left(\rho^{2}\left(0\right)\right),
\EqLabel{eq:SQRE}
\end{align}
with a constant $\beta$, \ie, $\beta_{l}=\beta$. We do not assume that $\beta$ is increasing or decreasing, or that $\beta_{\infty}=\infty$ or $\beta_{1}=\infty$. We do not believe that it is more difficult to perform more iterations because all iteration processes are supposed to be fictitious. We will demonstrate that essentially none of the desired features of SQRE rely on details concerning orders of $\beta$ or increasing or decreasing values of $\beta$, instead depending on iteration. The iteration alone suffices to lead us to the preferred NE. Furthermore, we have not found a thorough discussion of the stability of NRS solutions. Thus, this manuscript can be regarded as a further development of the NRS in that it distinguishes stable from unstable solutions and notes that the key component is the iteration and not the order of $\beta$ or the assumption of the limit of $\beta_{n}$ approaching $\infty$.

In sum, the proposed interative process differs from many other theories in that it is a map from a mixed strategy of all of the players to a new mixed strategy of all of the players. One might have some questions with respect to the interpretation of this process and comparison with real game playing. However, in physics, it is natural to study the evolution of distribution functions directly instead of the evolution of individual trajectories. Additionally, we do not aim at making the process reflect to realty more closely, but only to make the long-term solution more capable of predicting the game outcomes. Next, we discuss certain features of the proposed iterative process and compare its solutions to observed behavioral outcomes of experiments.

\section{Major features of ILQRD, illustrated through examples}
\label{sec:example}

In this section, first, we demonstrate by example that the QRE covers all NEs, including pure and mixed NEs. This conclusion is not new and has been implicitly demonstrated in \cite{McKelvey:QRE_NE}. For a $2\times 2$ game, this statement can be proved by considering \Eqref{eq:iteration21} and \Eqref{eq:iteration12} in the extreme case of $\beta\rightarrow \infty$. We present a general proof in section $\S$\ref{sec:proof}. Second, we demonstrate that once there is a preferred pure NE in a game, our SQRE converges to this focal NE. This phenomenon serves as a natural refinement. Unfortunately, we have not proved this conclusion mathematically. Thus, we illustrate this outcome by examples. Third, we demonstrate that all QREs that correspond to mixed NEs are not stable for the case of large $\beta$ ($\beta\rightarrow \infty$) but that some of these QREs can be stable for finite $\beta$. The third conclusion first questions the predictive power of QREs (when they correspond to mixed NEs) and then redeems the QRE as a possibly applicable solution when bounded rationality is considered. This conclusion also distinguishes stable QREs from unstable QREs, which enables examination of  the applicability of QREs to the explanation of experimental observations or real-life phenomena. That is, in principle, it is no longer true that QREs are strictly better than NEs. If the experimental data of a game are located in the region of unstable QREs, the QRE is not a practical solution concept for the game because unstable solutions are not reachable following the evolution. As far as we know, the last two conclusions (first, that SQREs converge to preferred NEs when there are such NEs and thus it represent a natural refinement of NEs and, second, QREs that correspond to mixed NEs become unstable for large enough $\beta$) are new. This also implies that mixed NE are not directly applicable, since they are in a sense always unstable in the limit of large $\beta$ (also according to best-response dynamics), unless these mixed NEs are close to SQREs. We believe this also improves our understanding of mixed NEs.

\subsection{Games with two pure NEs and a mixed NE: Coordination Game and Hawk-Dove Game}
It can be demonstrated that for games with a dominant strategy, such as the prisoner's dilemma, the dominant strategy is such that one of the QREs and the SQRE converge toward the dominant strategy in the large $\beta$ limit. However, this case is trivial. We demonstrate behaviors of our ILQRD starting from the more interesting coordination game, which does not have any dorminant strategies. The payoff matrices of coordination game are as follows:
\begin{align}
G^{1,2} = \left[\begin{array}{cc} 1, 1 & 0, 0 \\ 0, 0 & 5, 5\end{array}\right].
\EqLabel{eq:papoff_CG}
\end{align}
It is known that there are two pure-strategy NEs and a mixed NE. They are $\left(p^{1}, p^{2}\right)=\left(0,0\right), \left(1,1\right), \left(0.83,0.83\right)$, which are denoted respectively as $\mbox{PNE}_{00}, \mbox{PNE}_{11}$ and $\mbox{MNE}$. Conventionally, the preferred NE --- $\mbox{PNE}_{00}$ --- of this game can be found to be the focal NE through refinement \cite{Selton:Selection, Samuelson:Selection}. Evolutionary stability analysis \cite{Weibull:Evolution} indicates that the mixed NE is unstable. That is, when $p^{1}_{0}<p^{1}_{c}=0.83$ ($p^{1}_{0}>p^{1}_{c}$), the population converges to $\mbox{PNE}_{00}$ ($\mbox{PNE}_{11}$). Because the $\beta$ we introduced has no absolute meaning, in all of the manuscript's remaining calculations, we normalize each player's payoffs by their own maximum. For the payoffs provided in \Eqref{eq:papoff_CG}, the maximums are $5$ and $5$ for the first and second player, respectively.

\begin{figure}
\center\includegraphics[width=8cm]{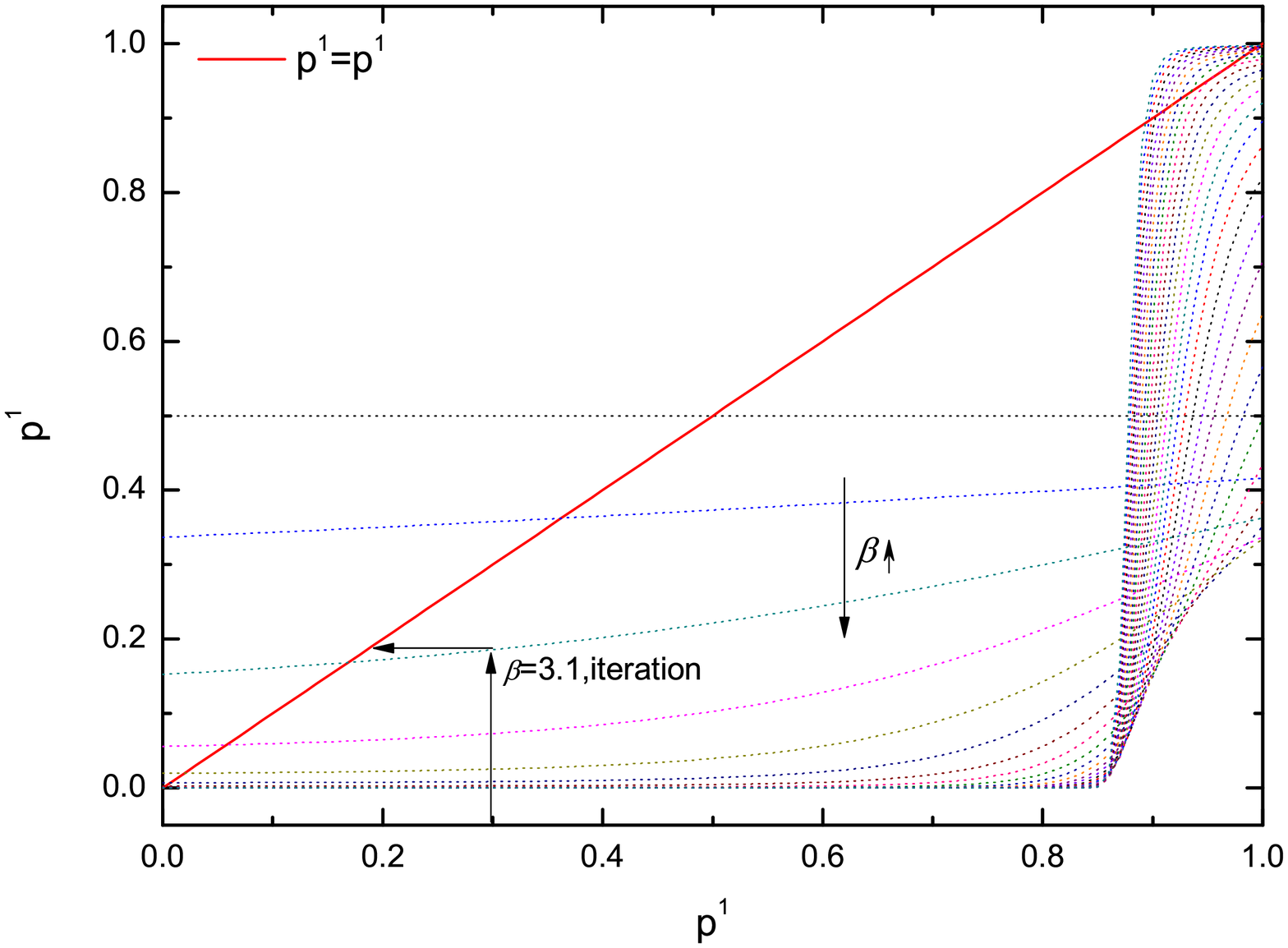}
\caption[Iterative mappings on CG]{The iterative mapping for various $\beta$ and the line of $p^{1}=p^{1}$ on the coordination game. For a given value of $\beta$, the intersections of corresponding curve and the line of $p^{1}=p^{1}$ are the QREs. For this game, there is only one QRE for small values of $\beta$, whereas there are three QREs($QRE_{00}$, $QRE_{11}$ and $QRE_{MNE}$) for larger $\beta$. Here, $QRE_{00}$ and $QRE_{11}$ are always stable, whereas $QRE_{MNE}$, which is close to $p^{1}=0.83$, is always unstable. As an example, we illustrate the first two steps of the iterative process for a specific value of $\beta=3.1$. }
\label{fig:mappingsCG}
\end{figure}

Fig.\ref{fig:mappingsCG} shows iterative mappings for a range of values of $\beta$ of this coordination game. Each curve except the red curve (which is $p^{1}=p^{1}$) is a curve of the iterative mapping for a given value of $\beta$. The long arrow labeled $\beta\uparrow$ indicates the shift of the curve of the iterative mapping when $\beta$ increases. As an example, we also illustrate the first two steps of the iterative process for a specific value of $\beta=3.1$. Usually it takes only less than $20$ iterations to find the SQREs with reasonable accuracy starting from any initial value of $p^{1}$. We can observe that for small $\beta$, there is only one QRE, which corresponds to $\mbox{PNE}_{00}$ (denoted as $QRE_{00}$). For large $\beta$, there are multiple QREs: $QRE_{00}$, a QRE that corresponds to $\mbox{PNE}_{11}$ (denoted as $QRE_{11}$) and a QRE that corresponds to $\mbox{MNE}$ ($QRE_{MNE}$). However, not all of these QREs are equally good. One can observe that for this game $QRE_{00}$ and $QRE_{11}$ are always stable, whereas $QRE_{MNE}$ is always unstable. Here, stability means that if the initial guess is not correct at the QRE, one iteration step will drive the value of $p^{1}$ closer to the QRE. Those QREs that are stable in this sense are referred to as SQRE. To simplify our terminology, we will refer to the QREs that correspond to pure (mixed) NEs in the limit of $\beta\rightarrow\infty$ as pure (mixed) QREs, although all QREs for all finite $\beta$ are in fact mixed strategies. The same naming scheme and the same notation are used for SQREs, for instance, pure $SQRE_{00}$, pure $SQRE_{11}$ and mixed $SQRE_{MNE}$.

\begin{figure}
\center\includegraphics[width=8cm]{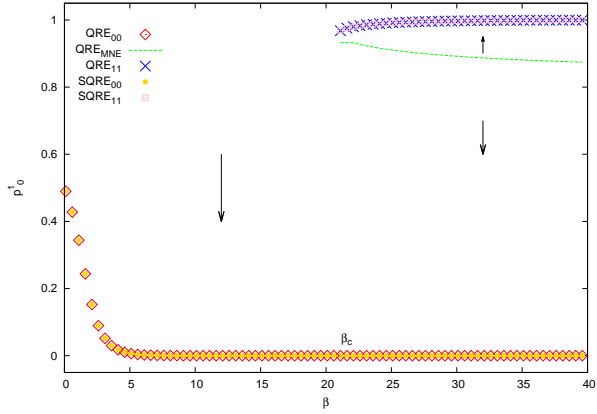}
\caption[Phase diagram on CG]{Dependence on $\beta$ of SQREs of the coordination game. When $\beta<\beta_{c}$, which here is approximately $22$, there is only one QRE($QRE_{00}$) and it is a SQRE. When $\beta>\beta_{c}$, there are three QREs. However, the $QRE_{MNE}$ is unstable. For initial values of $p^{1}_{0}$ less than the $p^{1}$ of $QRE_{MNE}$ (the green line), the iteration results in the $QRE_{00}$ (the pink square; thus, $SQRE_{00}$ in this case). Otherwise, the $QRE_{11}$ becomes the long-term solution of the iteration $SQRE_{11}$ (the gold circle). The corresponding $p^{2}$ (not shown) can be straightforwardly calculated.}
\label{fig:phaseCG}
\end{figure}

In principle, all of the information on the QREs and SQREs of this game is included in Fig. \ref{fig:mappingsCG}. However, to better illustrate the stability of QREs, in Fig. \ref{fig:phaseCG}, we plot the dependence on $\beta$ of the stability of the QREs of the coordination game . From the lower left section, where $QRE_{00}$ overlaps with $SQRE_{00}$, we can observe that for small $\beta$ ($\beta<\beta_{c}$, which here is approximately $22$), there is only one QRE, and it is a SQRE. For all initial values of $p^{1}$, this SQRE is the only long-term state. When $\beta>\beta_{c}$, there are three QREs: $QRE_{00}$, $QRE_{11}$ and $QRE_{MNE}$. However, $QRE_{MNE}$ is unstable. For initial values of $p^{1}_{0}$ less than the unstable QRE (the green line, $QRE_{MNE}$), the iteration results in $SQRE_{00}$(the pink line). Otherwise $SQRE_{11}$(the blue line) becomes the iteration's long-term solution. The corresponding $p^{2}$ (not shown in the figure) can be straightforwardly calculated using \Eqref{eq:iteration12}. Note from the upper right section that the region between $SQRE_{11}$ (the blue line) and $QRE_{MNE}$ (the green line) is narrow compared with the space between $SQRE_{00}$ (the blue line) and $QRE_{MNE}$ (the green line). This outcome indicates that for a wide range of initial value of $p^{1}$ the SQRE of this game converges toward $SQRE_{00}$, which is also $\mbox{PNE}_{00}$, the preferred NE in this game. This picture, particularly the right half of Fig. \ref{fig:phaseCG}, is highly similar to results from evolutionary stability analysis. However, the behavior with small $\beta$ such that no matter what the initial value of $p^{1}$ is the SQRE is always the QRE that corresponds to $\mbox{PNE}_{00}$, is a unique result of our ILQRD. We believe that this unique SQRE for small $\beta$ can be regarded as a refinement of NEs.

In this game, we observe that QREs cover all NEs and that for a wide range of initial values of $p^{1}$, $SQRE_{00}$ is the SQRE, and it corresponds to the preferred NE ($p^{1}=0$). Particularly when $\beta<\beta_{c}$ for all values of $p^{1}_{0}$, the preferred NE is the only SQRE. It is intuitive to expect that even for large $\beta$ the green line ($QRE_{MNE}$) is closer to the blue line ($QRE_{11}$) than the pink line ($QRE_{00}$). Thus, for a wide range of $p^{1}_{0}$, $SQRE_{00}$ will be the long-term solution of the iterative process. In this sense, ILQRD and its SQRE represent a natural refinement for QREs and NEs. Additionally, we note that for this game, overall, our prediction is somewhat similar to that of evolutionary stability analysis. Next, we discuss another game, in which our prediction differs more significantly from the results of evolutionary stability analysis than the situation in this Coordination Game.

\begin{figure}
\center\includegraphics[width=8cm]{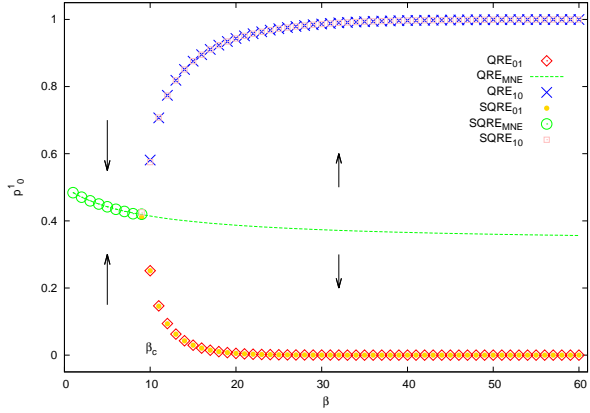}
\caption[Phase diagram of HD]{QREs and SQRE of the hawk-dove Game.  When $\beta<\beta_{c}$, which is approximately $10$ for this game, starting from an arbitrary initial value of $p^{1}$,  the long-term solution of the ILQRD of the hawk-dove Game is $SQRE_{MNE}$. When $\beta>\beta_{c}$, there are three QREs: $QRE_{01}$, $QRE_{10}$ and $QRE_{MNE}$. However, here, $QRE_{MNE}$ is unstable. In this case, the SQRE depends on the initial values of $p^{1}$: When it is above $p^{1}$ of $SQRE_{MNE}$ (the green line), it is $SQRE_{10}$ (pink square). Otherwise, it is $SQRE_{01}$ (the gold circle). }
\label{fig:PhaseHD}
\end{figure}

Now, we consider the hawk-dove game, which according to evolutionary game analysis \cite{Weibull:Evolution} has one evolutionary stable mixed NE and two evolutionary unstable pure NEs. Its payoff matrices are defined as follows:
\begin{align}
G^{1,2} = \left[\begin{array}{cc} 0, 0 & 7, 2 \\ 2, 7 & 6, 6\end{array}\right].
\end{align}
Based on a calculation of iterative mappings similar to that shown in Fig. \ref{fig:mappingsCG}, the QREs and SQREs of the hawk-dove game for various values of $\beta$ are plotted in Fig. \ref{fig:PhaseHD}. We observe that for small $\beta$, there is only one QRE($QRE_{MNE}$) and it is a SQRE. This SQRE is the long-term solution of the ILQRD for all of the initial values of $p^{1}$. For large $\beta$, there are three QREs: $QRE_{01}$, $QRE_{10}$ and $QRE_{MNE}$. However, in this case, $QRE_{MNE}$ is unstable. The long-term solution of the ILQRD depends on the initial values of $p^{1}$. When the initial values is above $p^{1}$ of $QRE_{MNE}$ (the green line), $QRE_{10}$ is the SQRE. Otherwise, $QRE_{01}$ is the SQRE. This figure provides more information than the QREs and NEs by distinguishing stable QREs from unstable ones. Additionally, this figur differs from the results of the evolutionary stability analysis, which demonstrates that the mixed NE of this game is evolutionary stable. Our results suggest that it is stable only when $\beta<\beta_{c}$, which is approximately $10$ for this game. Otherwise, the outcome of this game will prefer to be $QRE_{01}$ or $QRE_{10}$. This outcome differs from the results of evolutionary game analysis of the same game. For small $\beta$, the $QRE_{MNE}$ is the only SQRE, whereas for large $\beta$, the mixed QRE becomes unstable: depending on the initial value of $p^{1}$, different SQREs can be reached.

\subsection{Games with a unique mixed NE: Tennis Game}
The third example is the tennis game \cite{Dixit:GameTheory}, which has one mixed NE $\left(p^{1}, p^{2}\right)=\left(0.7,0.6\right)$ but no pure NEs. The payoff matrices are given as follows,
\begin{align}
G^{1,2} = \left[\begin{array}{cc} 5, 5 & 8, 2 \\ 9, 1 & 2, 8\end{array}\right].
\end{align}

\begin{figure}
\includegraphics[width=5.5cm]{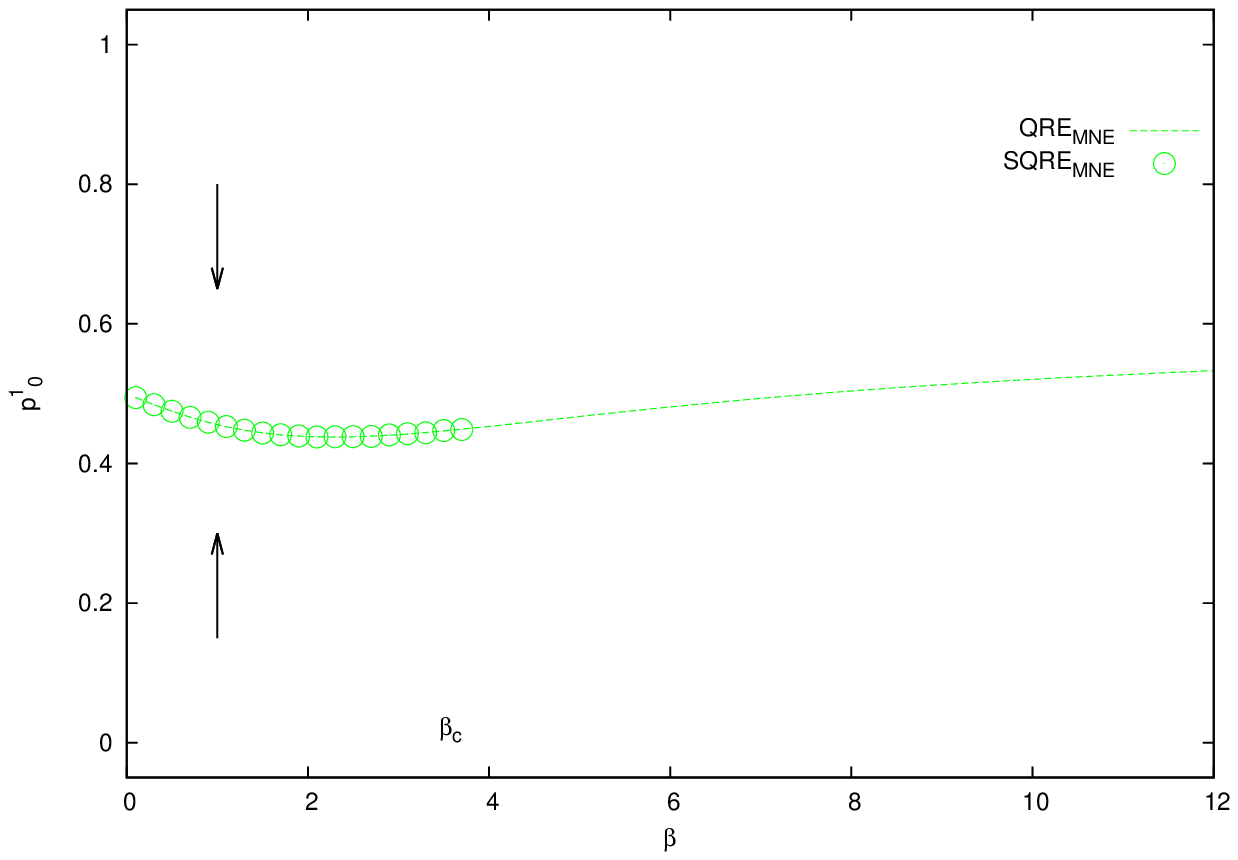} \includegraphics[width=7cm]{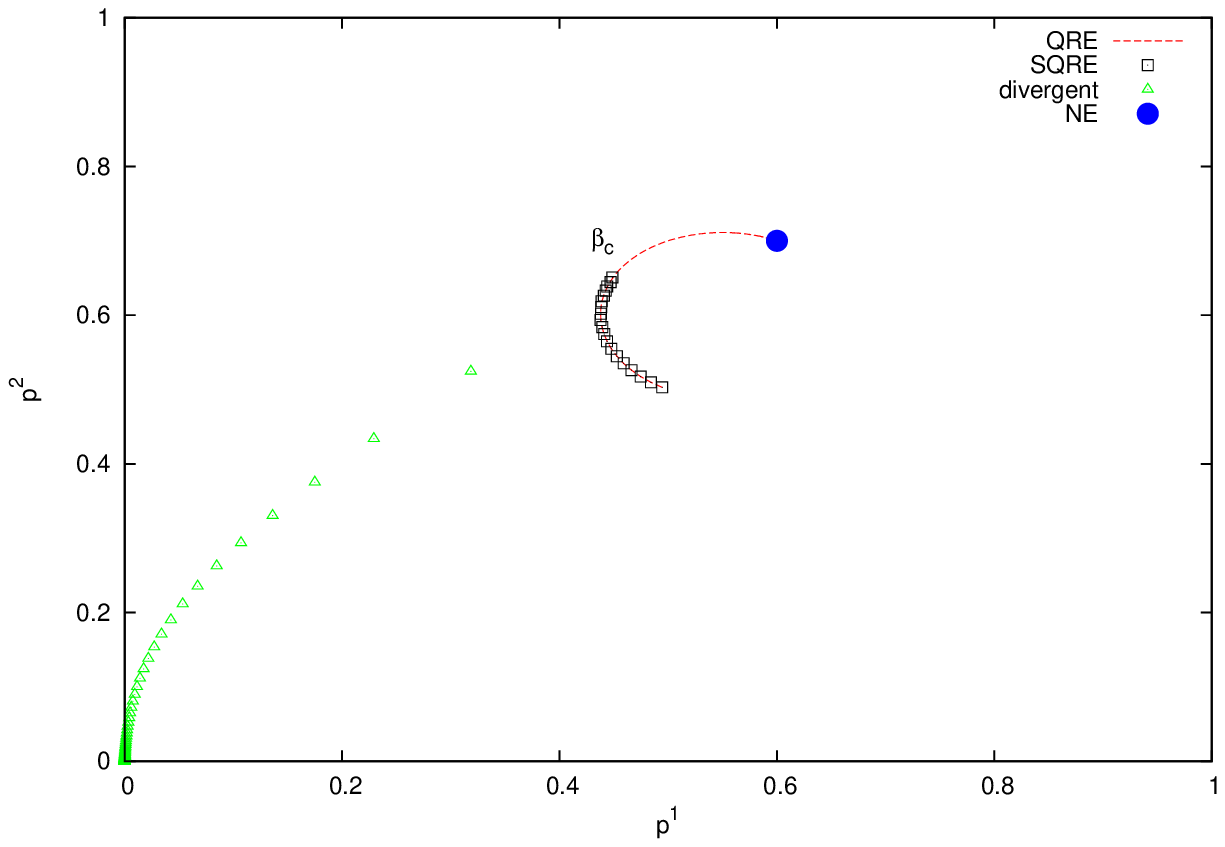}
\caption[NEs, QRE and SQRE of the Tennis Game]{NEs, QRE and SQRE of the tennis game. From (a), we observe that there is only one QRE ($QRE_{MNE}$: the green line) for each given value of $\beta$. There is one SQRE ($SQRE_{MNE}$: the green circle) when $\beta<\beta_{c}$, which is approximately $3.7$ for this game, and there is no SQRE for $\beta>\beta_{c}$. (b) shows the plots of $p^{1}$ and $p^{2}$ and their relation with $\beta$. The green curve of triangles  represents values of $\left(p^{1}, p^{2}\right)$ when such divergent iterations are performed $1000$ times. We observe again that when $\beta>\beta_{c}$, $QRE_{MNE}$ can not be reached by any SQREs, although it converges towards the NE.}
\label{fig:TG}
\end{figure}

Both the QRE and SQRE are shown in Fig. \ref{fig:TG}. We find that for this game there is always one and only one QRE for a given value of $\beta$ and that this QRE converges toward the mixed NE $\left(0.7,0.6\right)$ in the limit of $\beta\rightarrow\infty$. Therefore, this QRE is $QRE_{MNE}$. However, the SQRE follows this $QRE_{MNE}$ only when $\beta$ is small enough ($\beta \le \beta_c$, which for this game is approximately  $3.7$). When $\beta > \beta_c$, this $QRE_{MNE}$ becomes unstable; thus, there is no longer any SQRE. This game displays a substantial difference between the SQRE and the mixed NEs. Such games can be used to test the applicability of SQRE relative to NE, as shown in the following section.

From these examples, we observed the following: first, QREs exist for all of the games discussed above, and QREs cover all of the NEs in the limit of $\beta\rightarrow\infty$; second, for games with a preferred NE, the SQRE can be used as a refinement of the NEs; and finally,  mixed QREs become unstable for large enough $\beta$ values; thus, the SQRE can be regarded as a refinement of the QREs (and therefore the NEs). These observations reflect the three main features of our ILQRD. As we pointed earlier, the first feature was implicitly demonstrated in \cite{McKelvey:QRE_NE}. However, the latter two features are new. For certain games, the distance between the SQRE and the mixed NE is large. Thus, experimental results on such games are suitable for testing the applicability of the SQRE relative to the NE. Before we proceed to a comparison of our theoretical prediction with experimental results, we prove the previously mentioned first and the third features of our ILQRD. Unfortunately, we cannot prove the second feature because at present we do not know the necessary and sufficient condition for a game to have a preferred NE. Instead, we simply desire to demonstrate that for the coordination game and the hawk-dove game, for small $\beta$, the SQRE corresponds to a certain refinement: $SQRE_{00}$ for the former game\cite{Selton:Selection, Samuelson:Selection} and $SQRE_{MNE}$ for the latter \cite{Weibull:Evolution}.

\section{Proof of main conclusions on $2\times2$ symmetric games}
\label{sec:proof}

In this section, while considering a symmetric game for simplicity, we wish to prove the previously described three features using examples.

Let $a^{'}=a, b^{'}=b, c^{'}=c, d^{'}=d$ in \Eqref{eq:map2to1} and \Eqref{eq:map1to2}. If we merge the two equation and focus only on $p^{1}$, the iteration function becomes
\begin{align}
p^{1}\left(t+1\right)=\frac{1}{1+e^{\beta\left[\left(c-d-a+b\right)\left(\frac{1}{1+e^{\beta\left[\left(c-d-a+b\right)p^{1}\left(t\right)+d-b\right]}}\right)+d-b\right]}}\triangleq f\left(p^{1}\left(t\right);  \beta\right).
\EqLabel{eq:22Sgame}
\end{align}
First, there are five possible NEs: the pure NEs $\left(0,0\right), \left(1,0\right), \left(0,1\right), \left(1,1\right)$ and a mixed NE $\left(\frac{b-d}{c-d-a+b},\frac{c-d}{c-d-a+b}\right)$, depending on the values of $a, b, c$ and $d$. Let us first demonstrate that the QREs cover all of the NEs under proper conditional relations among $a, b, c$ and $d$. That is \Eqref{eq:22Sgame} has five possible solutions: $QRE_{00}\left(\beta\right)$, $QRE_{10}\left(\beta\right)$, $QRE_{01}\left(\beta\right)$,  $QRE_{11}\left(\beta\right)$ and $QRE_{MNE}\left(\beta\right)$, which corresponds to the previously mentioned five NEs in the limit of $\beta\rightarrow\infty$. In terms of these notations, we wish to demonstrate that
\begin{align}
\begin{cases}
QRE_{00}\left(\beta\right)  & \mbox{if } b-d<0 \mbox{ \& } c-a>0 \\
QRE_{00}\left(\beta\right) \mbox{ \& } QRE_{11}\left(\beta\right)  & \mbox{if } b-d<0 \mbox{ \& } c-a<0 \\
QRE_{10}\left(\beta\right)  \mbox{ \& } QRE_{01}\left(\beta\right)  & \mbox{if } b-d>0\mbox{ \& } c-a>0 \\
QRE_{11}\left(\beta\right)  & \mbox{if } b-d>0\mbox{ \& } c-a<0 \\
QRE_{MNE}\left(\beta\right) & \mbox{if } 0< \frac{b-d}{b-d+c-a} < 1
\end{cases} .
\end{align}
We will prove that the first and the last cases and the extensions to other cases are trivial.

It is straightforward to demonstrate that when $b-d<0 \mbox{ \& } c-a>0$,
\begin{align}
f\left(0; \beta\right)>0 \mbox{ and } \lim_{\beta\rightarrow \infty}f\left(0; \beta\right)=0 \mbox{ and } 0<\lim_{\beta\rightarrow \infty}\frac{\partial f\left(0; \beta\right)}{\partial p}<1.
\EqLabel{eq:case1}
\end{align}
These three conditions mean that $f(0;\beta)$ is always greater than $0$. However, it approaches more closly to $0$ when $\beta$ increases. Additionally, when $\beta$ is sufficiently large, $f(p;\beta)$ increases near $p=0$. However, it increases slower than $p^{1}$. This statement is equivalent to stating that $f\left(0; \beta\right)-0>0$, and there is a $\bar{p}$ such that $f\left(\bar{p}; \beta\right)-\bar{p}<0$ when $\beta$ is sufficiently large. Thus, there must be a $p^{*}\left(\beta\right)$, such that $f\left(p^{*}; \beta\right)-p^{*}=0$. In the limit of $\beta\rightarrow \infty$ such $p^{*}\left(\beta\right)=0$. The situation for $p^{2}$ can be analyzed similarly.

For $QRE_{MNE}$, where  $0<p^{*}=\frac{b-d}{b-d+c-a} <1$  is a proper mixed strategy, we first want to demonstrate that
\begin{align}
 \lim_{\beta\rightarrow \infty}f\left(p^{*}; \beta\right)=p^{*} \mbox{ and }  \lim_{\beta\rightarrow \infty}\left.\frac{\partial f\left(p; \beta\right)}{\partial p}\right|_{p^{*}}-1\neq 0.
\EqLabel{eq:pMNE}
\end{align}
If we can prove this point, then, first, $p^{*}$ is a fixed point in the limit of $\beta\rightarrow \infty$. Second, this fixed point is not a maximum or minimum of the function $f\left(p; \beta\right)-p$. The latter means that the curve $f\left(p; \beta\right)-p$ passes across $0$ when $\beta$ is sufficiently large and $0$ is not an extremum.

Because we are working with symmetric games, the iteration function $f\left(p; \beta\right)$ can be regarded as a composite mapping of the following function:
 \begin{align}
f\left(p; \beta\right)=g\left(g\left(p; \beta\right); \beta\right) = g \circ g,
\EqLabel{eq:composite}
\end{align}
where
\begin{align}
g\left(p; \beta\right)=\frac{1}{1+e^{\beta\left[\left(c-d-a+b\right)p\left(t\right)+d-b\right]}}.
\EqLabel{eq:g}
\end{align}
In terms of this function, we can then easily demonstrate that
\begin{align}
 \lim_{\beta\rightarrow \infty}g\left(p^{*}; \beta\right)=p^{*} \mbox{ and }  \lim_{\beta\rightarrow \infty}\left.\frac{\partial g\left(p; \beta\right)}{\partial p}\right|_{p^{*}}-1\neq 0.
\EqLabel{eq:gMNE}
\end{align}
Here, we discuss the first part in greater detail, whereas the remainder is straightforward. The fixed point of the mapping $g$ is defined equivalently as follows:
\begin{align}
p=\frac{\frac{1}{\beta}\ln\left(\frac{1}{p}-1\right)+b-d}{c-d-a+b}
\EqLabel{eq:pexact}
\end{align}
For $p\left(t\right)$ that is close enough to $p^{*}$ such that $0<p\left(t\right)<1$, the limit of the RHS of \Eqref{eq:pexact} when $\beta\rightarrow\infty$ becomes exactly $p^{*}$.

Up to this point, we have demonstrated that the QREs of \Eqref{eq:22Sgame} cover all of the possible pure strategy NEs and the mixed NEs. Next, we discuss their stability. We wish to demonstrate that any pure QREs if exist are always SQREs and that mixed QREs are SQREs only for small $\beta$ but unstable for large $\beta$. Additionally, we attempt to define $\beta_{c}$, the critical value of $\beta$.

Stable solutions are a subset of fixed points. For the iterative mapping defined in \Eqref{eq:22Sgame}, we use linear stability analysis    \cite{Strogatz:LinearStability}, according to which a solution of \Eqref{eq:22Sgame} is stable if the Jacobian (simply a derivative in this case) of the right-hand side is less than $1$, \ie,
\begin{align}
\left.\frac{\partial f}{\partial p}\right|_{p^{*}}=\left[\beta\left(c-d-a+b\right)\right]^2\left(1-p^{*}\left(\beta\right)\right)^2\left(p^{*}\left(\beta\right)\right)^2<1,
\EqLabel{eq:stable}
\end{align}
where $p^{*}\left(\beta\right)$ is defined as in
\begin{align}
p^{*}=f\left(p^{*}; \beta\right).
\EqLabel{eq:fixedpoint}
\end{align}

For pure QREs, for example, $QRE_{00}$, $p^{*}\left(\beta\right)$ decreases to $0$ exponentially as $\frac{1}{1+e^{\beta\left(d-b\right)}}$. Therefore, $\beta p^{*}\left(\beta\right) \rightarrow 0$. Thus, \Eqref{eq:stable} is always satisfied. Other pure QREs can similarly be shown to be stable. Thus, all of them are SQREs.

\begin{figure}
\center\includegraphics[width=8cm]{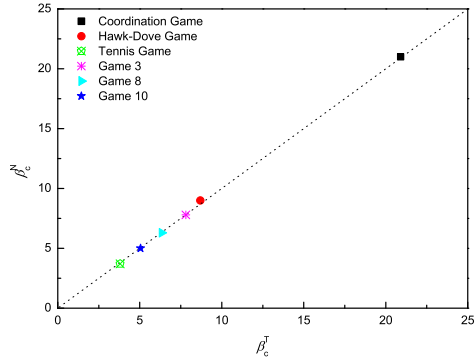}
\caption[$\beta^{N}_{c}$ v.s. $\beta^{T}_{c}$ for all discussed games]{For all of the games discussed above and below $\beta^{N}_{c}$, which is the $\beta_{c}$ found in iterative calculation, is plotted against $\beta^{T}_{c}$, the $\beta_{c}$ which is solved.}
\label{fig:betac}
\end{figure}

For mixed QREs, first consider the case of $\beta=0$. In this case, the payoff makes no difference. therefore, $p^{*}\left(\beta=0\right)=\frac{1}{2}$, and thus
\begin{align}
\left.\frac{\partial f}{\partial p}\right|_{p^{*}}=0<1,
\end{align}
Therefore, mixed QREs are SQREs in the limit of $\beta\rightarrow 0$. Now, consider the case of $\beta\rightarrow \infty$. Because $0<p^{*}\left(\beta\right)<1$ is a finite number, there is always a $\beta_{c}$ such that when $\beta>\beta_{c}$, \Eqref{eq:stable} is no longer valid. $\beta_{c}$, which is the critical value of $\beta$, is defined as follows:
\begin{align}
\left[\beta_{c}\left(c-d-a+b\right)\right]^2\left(1-p^{*}\left(\beta_{c}\right)\right)^2\left(p^{*}\left(\beta_{c}\right)\right)^2=1,
\EqLabel{eq:betac}
\end{align}
For a given game with fixed values of $a, b, c$ and $d$, the numerical value of $\beta_{c}$ can be solved numerically using \Eqref{eq:fixedpoint} and \Eqref{eq:betac} together. In Fig. \ref{fig:betac}, we plot all $\beta_{c}$ found in numerical simulation of ILQRD (denoted as $\beta^{N}_{c}$) against the $\beta_{c}$ solved from \Eqref{eq:fixedpoint} and \Eqref{eq:betac} (denoted as $\beta^{T}_{c}$). We found that the two values agree with each other well.

For asymmetric games, similar $\beta_{c}$ can be derived. For simplicity, let us define
$C\left(p^{*}, \beta_{c};a,b,c,d\right) \equiv \beta_{c}\left(c-d-a+b\right)\left(1-p^{*}\left(\beta_{c}\right)\right)p^{*}\left(\beta_{c}\right)$. Then, the critical value of $\beta$ is determined by
\begin{align}
\left|C\left(p^{1,*}, \beta_{c};a,b,c,d\right)C\left(p^{2,*}, \beta_{c};a^{\prime},b^{\prime},c^{\prime},d^{\prime}\right)\right|=1, \\
p^{1,*}=g\left(p^{2,*}; \beta_{c} \left|\right. a, b, c, d\right), \\
p^{2,*}=g\left(p^{1,*}; \beta_{c} \left|\right. a^{\prime}, b^{\prime}, c^{\prime}, d^{\prime}\right).
\EqLabel{eq:betac2}
\end{align}
 for all the examples used in the previous section and also those payoff matrices of the experiments discussed in the next section, we plot $\beta_{c}$, which are numerically solved from these equations and the corresponding ones found from simulations. We found that values of $\beta_{c}$ numerically solved and values found from simulations are in very good agreement.

We have demonstrated that for symmetric $2\times2$ games, our ILQRD has the following characteristics: $(1)$ the QREs cover all of the pure and mixed NEs, and $(2)$ pure QREs are SQREs, whereas mixed QREs are SQREs for small $\beta$ but unstable for sufficiently large $\beta$. These conclusions cover all the major features that we demonstrated in the preceding section using specific examples. A general proof for general $2\times2$ games should be straightforward, although it would involve more tedious algebra.

From this general proof, we have observe that for all of the values of $\beta$, the QRE exists but not the SQRE. Intuitively, when $\beta$ increases, the QREs approach NEs because our iteration process moves closer to the best response dynamics. However, QREs might lose their stability when $\beta$ is sufficiently large. The difference between the SQRE and the NE depends on the competition between these two effects: approaching NE and losing stability. This difference provides a means of examining the applicability of the QRE. The QRE has been criticized because the fixed points, which are what we refer to as QREs, can always surpass mixed the NE  as a result of the free parameter $\beta$. We agree with this statement. Therefore, that experimental data are closer to the QREs than the mixed NEs does not imply that the QRE is a better solution concept than the NE. However, this criticism does not hold for our SQRE, which loses its stability for larger $\beta$, indicating that our SQRE cannot always outperform mixed NEs. By assessing whether the experimental data points are closer to our SQRE than the mixed NEs, we can compare the SQRE solution concept with the NE(section $\S$\ref{sec:exp}).

\section{Difference between QRSS and SQRE}
\label{sec:QRSS}

QRSS, which is occasionally referred to as Logit response dynamics \cite{Alos:Logit, Konno:Logit_Exact}, starts from an arbitrary strategy profile for each player and then uses the transition matrix defined in \Eqref{eq:QRSS} to evolve the strategic states of all of the players into the invariant distribution defined in \Eqref{eq:QRSSFinal}. Its stationary states have been discussed by several researchers. However, there is no full study on the necessary and sufficient conditions of the convergence of the QRSS to NE  \cite{Alos:Logit, Baron:QRSS, Konno:Logit_Exact}. In this section, we demonstrate that although \Eqref{eq:QRSS} appears similar to \Eqref{eq:QRE}, $P_{ss}$ as defined in \Eqref{eq:QRSSFinal} differs substantially from $\rho_{\infty}\left(\beta\right)$. As discussed below, the difference is that $P_{ss}$ is a distribution in $\Delta\left(S\right)$ which is the set of all of the possible distribution functions on $S$, whereas $\rho_{\infty}\left(\beta\right)$ is a member of $\Delta$, which is the set of independent distribution functions. That is, $P_{ss}$ includes possibly correlated strategies, whereas $\rho_{\infty}\left(\beta\right)$ describes only purely non-cooperative strategies.

\begin{figure}
\center\includegraphics[width=8cm]{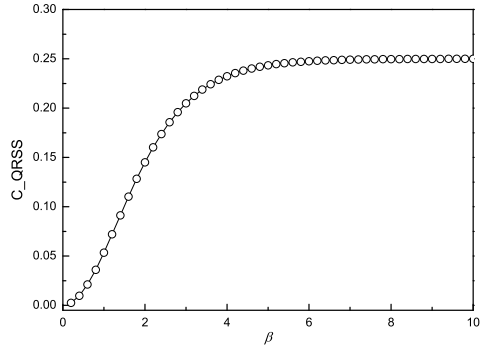}
\caption[Difference between QRSS and SQRE]{Correlation index $C$ obtained from QRSS plotted against $\beta$. The correlation index of the QRSS is always non-zero except, again, in cases of extremely small $\beta$.}
\label{corelation}
\end{figure}

Now, we demonstrate this statement using one example: a $2\times2$ symmetric game with payoff matrices  \cite{Alos:Logit}:
\begin{align}
G^{1,2} = \left[\begin{array}{cc} 1,1 & 0, 0 \\ 0, 0 & 1, 1\end{array}\right].
\end{align}
This game is a potential game  \cite{Alos:Logit}. It has two strict NEs and a proper mixed NE. According to   \cite{Alos:Logit} and   \cite{Konno:Logit_Exact}, the corresponding Logit response dynamics has an invariant distribution of strategy profiles that correspond to the potential maximizer and the proper mixed NE. Here, we show that the invariant distribution in fact does not correspond to the proper mixed NE. It involves correlations between players. First, we calculate the probability transition matrix $M$ and then obtain the $P_{ss}$ of this game according \Eqref{eq:QRSSFinal}. From the $P_{ss}$, we find the reduced strategy profile $P^{1}$ for player $1$ and $P^{2}$ for player $2$:
\begin{align}
P^{1}\left(s^{1}_{l}\right) = \sum_{m}P_{ss}\left(s^{1}_{l}, s^{2}_{m}\right), P^{2}\left(s^{2}_{m}\right) = \sum_{l}P_{ss}\left(s^{1}_{l}, s^{2}_{m}\right).
\end{align}
Then, we define the correlation index as
\begin{align}
C= \sum_{l,m}\left[P_{ss}\left(s^{1}_{l}, s^{2}_{m}\right)-P^{1}\left(s^{1}_{l}\right)P^{2}\left(s^{2}_{m}\right)\right]^2.
\end{align}
The correlation index is zero when the joint distribution $P_{ss}$ is a product of two independent probability distributions. Otherwise, it is nonzero and vice versa. In Fig.\ref{corelation}, we plot the correlation index $C$ calculated for the QRSS. We find that the correlation index of the QRSS is always non-zero except in cases of extremely small $\beta$ values. It is obvious that the SQRE is uncorrelated and that the correlation index remains $0$. In fact, the joint distribution is defined for the SQRE in this manner.

In this paper, we have not yet discussed the applicability of correlated strategies in non-cooperative games  \cite{Aumann:CE} and will not address it. However, we have noted that the invariant distribution of the QRSS or Logit response dynamics generally result in correlated strategies and the QRSS differs from the QRE and the SQRE.

\section{Experimental results re-analyzed using ILQRD}
\label{sec:exp}

As previously explained, a key difference between this manuscript and other studies on the QRE is that beyond certain values of $\beta$, QRE becomes unstable. Therefore, even when the experimental data are better fit by the QRE, if the data is near the unstable region, then applicability of the QRE is questionable. In this section, we examine how close the experimental data are to SQRE. We focus on experiments that involve $2 \times 2$ games with unique mixed NE and require the games to have a SQRE relatively distant from such mixed NE, \ie, games such as the one in Fig.\ref{fig:TG}. Erev et al.  \cite{Roth:data} conducted forty $2 \times 2$ constant sum games. Each pair of players played one game 500 times. Among these experimental games, there were ten games in which each game was played by nine pairs of subjects, whereas the other thirty games were played by one pair of subjects. Here, we use only experimental data from the ten games played by nine subject pairs. The payoff matrices of the ten games are shown in Table \ref{payoff}, which is reproduced from Table 1 in   \cite{Roth:data}. The sixth column shows the NE of each game. Each player was asked to choose between A and B. The payoff entry $AB$ presents player $1$'s wining probability ($\times 100$) when the player choose $A$ and the player's opponent choose $B$ and so on. The payoff for each win was $4$ cents. All ten games were played by fixed pairs for $500$ trials. Table \ref{data} shows the proportion of A choices in the $500$ trials by each player. We would have preferred the data from the $500$ pairs of independent players to the data from the games repeated $500$ times by the same pair of players. However, first, we did not find such data, and second, for this constant-sum game, it is believed that a repeated game produces no surprising result. For social dilemma games, such as the prisoner's dilemma, of which when the game is repeated even a finite number of times, the experimental behavioral outcome completely changes.

\begin{table}[htbp]
\caption {Payoff matrix of ten game samples}
\label{payoff}
\centering
\begin{tabular*}{0.8\textwidth}{@{\extracolsep{\fill}}cccccc}
\hline
Game & AA& AB & BA & BB & NE $(p^1, p^2)$\\
\hline
1 & 77 & 35 & 8 & 48 & (0.4878,0.1585)\\
2 & 73 & 74 & 87 & 20 &(0.9853,0.7941)\\
3 & 63 & 8 & 1 & 17 & (0.2253,0.1268)\\
4 & 55 & 75 & 73 & 60 & (0.3939,0.4545) \\
5 & 5 & 64 & 93 & 40 & (0.4732,0.2143)\\
6 & 46 & 54 & 61 & 23 & (0.8261,0.6739)\\
7 & 89 & 53 & 82 & 92 & (0.2174,0.8478)\\
8 & 88 & 38 & 40 & 55 & (0.2308,0.2615)\\
9 & 40 & 76 & 91 & 23 & (0.6538,0.5096)\\
10 & 69 & 5 & 13 & 33 & (0.2381,0.3333)\\
\hline
\end{tabular*}
\end{table}

\begin{table}[htbp]
\caption {The proportion of A choices in 500 trials by each of the players.}
\label{data}
\centering
\begin{tabular*}{0.6\textwidth}{@{\extracolsep{\fill}}ccc}
\hline
Game & $p^1$ & $p^2$ \\
\hline
1 & 0.591 & 0.318 \\
2 & 0.84 & 0.36 \\
3 & 0.583 & 0.222 \\
4 & 0.274 & 0.502 \\
5 & 0.378 & 0.32 \\
6 & 0.638 & 0.41 \\
7 & 0.295 & 0.522 \\
8 & 0.4 & 0.226 \\
9 & 0.562 & 0.449 \\
10 & 0.32 & 0.202 \\
\hline
\end{tabular*}
\end{table}

\begin{figure}
\includegraphics[width=8cm]{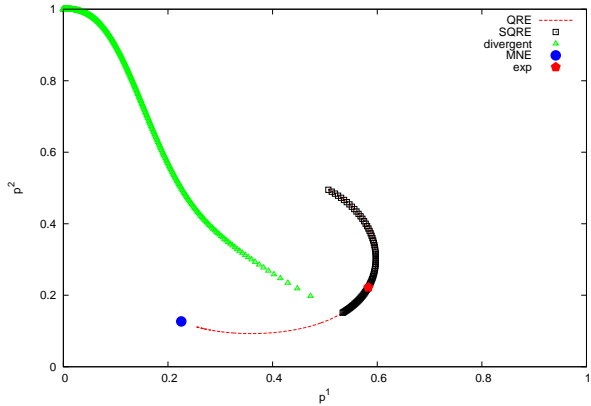}
\caption{NE, QRE, SQRE and experimental data of game $3$: the average strategy profile is in the region of the SQRE of this game and relatively distant from the NE. Five games out of the ten games exhibit a similar behavior. In the inset is the payoff matrix of this game. }
\label{data2}
\end{figure}

\begin{figure}
\includegraphics[width=8cm]{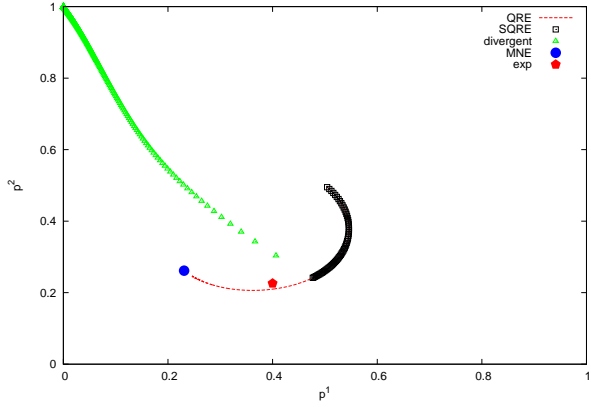}
\caption{NE, QRE, SQRE and experimental data of game $8$: the average strategy profile is in the region of unstable solutions but remains closer to the SQRE than to the NE. Three games out of the ten games exhibit a similar behavior. In the inset is the payoff matrix of this game.}
\label{data8}
\end{figure}

\begin{figure}
\includegraphics[width=8cm]{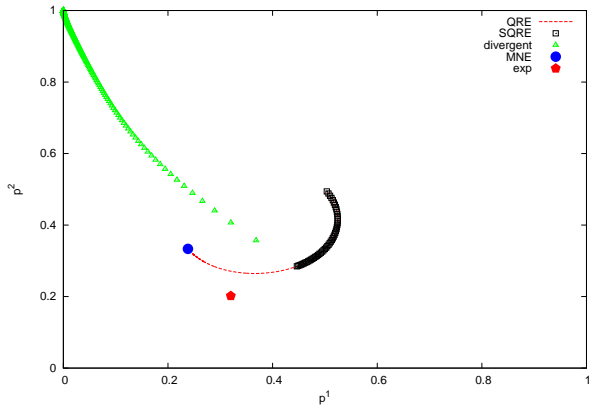}
\caption{NE, QRE, SQRE and experimental data of game $10$: the average strategy profile is closer to the NE than to the SQRE. Two games out of the ten games exhibit a similar behavior. In the inset is the payoff matrix of this game.}
\label{data10}
\end{figure}

According to the payoff matrices in Table \ref{payoff}, we obtain the SQRE and the QRE of ten games and compare the experimental data points with our SQRE. There are five games (games $2,3,6,7,9$) for which the experimental data points are in the area of the SQRE. We show one of them in Fig.\ref{data2} as an example. The experimental data from three games (games $1,5,8$) are in the area of unstable solutions but remain closer to our SQRE than to the NE(Fig.\ref{data8}). As shown in Fig.\ref{data10}, the experimental data points of the remaining two games (games$4,10$) are closer to the NE than to the SQRE. From these simple and limited comparisons, we conclude that the SQRE fits the observed behavior in real experiments better than the NE ($5+3:2$ in favor of the former in this limited comparison). However, further examination of the relation between the experimental data and our SQRE is required to arrive at a more definitive answer. We do not yet have any qualitative or quantitative criteria of games whose expected experimental behavior is close to the SQRE. This question will be examined in future investigations.

\section{Conclusion and Discussion}
\label{sec:conclusion}

In this manuscript, using the Logit quantal response function form (the Boltzmann distribution in statistical physics) to link the choice of strategy to the corresponding payoff in every step, we construct an iterative Logit quantal response dynamic process. Thus, the manuscript can be regarded as a dynamic version of Logit quantal response equilibrium. Importantly, our dynamic process differs from the so-called Logit response dynamics, which generally results in correlated equilibrium, even for non-cooperative games.

It has been shown in \cite{McKelvey:QRE1} that the QRE exists for all of the values of $\beta$ -- a measure of level of players' payoff sensitivity -- and converges toward NEs when $\beta \rightarrow \infty$. It has also been demonstrated on some examples and been taken for sure by some researchers that in fitting experimental data, the QRE is generally better than the NE because it is free to change the value of $\beta$ to improve the fitting \cite{Haile:QRE2,Morgan:QRE_Experiment, McKelvey:QRE_Experiment}. In our manuscript, we demonstrate that this is not the case: When taking stability into consideration, in principle, the QRE is no longer always better than the NE. Based on the dynamic process, stable and unstable QREs are distinguished. We find the following: ($1$) For games with a single focal pure NE, there is always one stable QRE that converges toward the preferred NE when $\beta \rightarrow \infty$. ($2$) For games without any focal pure NEs but with one unique proper mixed NE, when the payoff sensitivity $\beta$ is sufficiently large ($\beta>\beta_{c}$), the QREs lose their stability and become unstable. For certain games, the QREs are already close to their corresponding NEs before they lose their stability. Therefore, the difference between stable QREs and NEs is small. For other games, the difference between stable QREs and NEs is substantially more pronounced.

The latter case could be used to assess the applicability of the QRE to experiments and real-life observations. Then, we compared the stable and unstable QREs with experimental data. We found that the experimental observation of certain games ($5$ games from our preliminary tests) yields results within the regions of the stable QRE, that in other games ($3$ games), the experimental data are located in the unstable regions but remain closer to stable QREs than to the mixed NEs and that for other games ($2$ games) the experimental results are closer to the mixed NEs than to the stable QREs. We also believe that Linking mixed NEs to  mixed SQREs improves our understanding of applicability of mixed NEs. 

We have not identified any qualitative or quantitative criteria with which to classify games from this perspective. Further experimental and theoretical investigations are required to reach such a conclusion. In section \ref{sec:proof}, we only present a proof of the main observed features of our dynamic process for symmetric $2\times2$ games. In the future, a general discussion of the features and their proof for $N\times M$ games should be undertaken and cross-game experiments when performed and compared against our SQREs with estimated value of $\beta$ of a fixed group of players are of good value to put our concepts of SQREs up to further eximinations.

\section{Acknowledgements}

The authors thank Zhijian Wang and Bin Xu for sending us their paper on game theory and the principle of maximum entropy  \cite{Xu:PLA} and for sharing experimental data from their research. It was in their study \cite{Xu:PLA} that we found the additional experimental results of \cite{Roth:data}.

This paper was supported by the Fundamental Research Funds for the Central Universities.

\bibliographystyle{unsrt}
\bibliography{game}
\end{document}